\documentclass[10pt,letterpaper]{article}
\usepackage[top=0.85in,left=2.75in,footskip=0.75in]{geometry}

\usepackage{amsmath,amssymb}

\usepackage{changepage}

\usepackage[utf8x]{inputenc}

\usepackage{textcomp,marvosym}

\usepackage{cite}

\usepackage{nameref,hyperref}

\usepackage[right]{lineno}

\usepackage{microtype}
\DisableLigatures[f]{encoding = *, family = * }

\usepackage[table]{xcolor}

\usepackage{array}

\newcolumntype{+}{!{\vrule width 2pt}}

\newlength\savedwidth

\raggedright
\usepackage[aboveskip=1pt,labelfont=bf,labelsep=period,justification=raggedright,singlelinecheck=off]{caption}

\makeatletter
\renewcommand{\@biblabel}[1]{\quad#1.}
\makeatother

\usepackage{lastpage,fancyhdr,graphicx}
\usepackage{epstopdf}
\fancyheadoffset[L]{2.25in}
\fancyfootoffset[L]{2.25in}
\DeclareMathOperator*{\argmax}{\displaystyle argmax}

\begin{document}
\vspace*{0.2in}

\begin{flushleft}
{\Large
\textbf\newline{Evolutionary model discovery of causal factors behind the socio-agricultural behavior of the ancestral Pueblo} 
}
\newline
\\
Chathika Gunaratne\textsuperscript{1},
Ivan Garibay\textsuperscript{1},
Nguyen Dang\textsuperscript{2}
\\
\bigskip
\textbf{1} Complex Adaptive Systems Lab, Department of Industrial Engineering and Management Systems, University of Central Florida, Orlando, Florida, USA
\\
\textbf{2} AI group, School of Computer Science, University of St Andrews, St Andrews, UK
\\
\bigskip

%
%





* Corressponding authors
\\
Email: chathika@knights.ucf.edu (CG) 
\\
Email: ivan.garibay@ucf.edu (IG)

\end{flushleft}
\section*{Abstract}
Agent-based modeling of artificial societies offers a platform to test human-interpretable, causal explanations of human behavior that generate society-scale phenomena. However, parameter calibration is insufficient to conduct an adequate data-driven exploration of the importance of causal factors that constitute agent rules, resulting in models with limited causal accuracy and robustness. We introduce evolutionary model discovery, a framework that combines genetic programming and random forest regression to evaluate the importance of a set of causal factors hypothesized to affect the individual's decision-making process. We investigated the farm plot seeking behavior of the ancestral Pueblo of the Long House Valley simulated in the Artificial Anasazi model our proposed framework. We evaluated the importance of causal factors not considered in the original model that we hypothesized to have affected the decision-making process. Contrary to the original model, where closeness was the sole factor driving farm plot selection, selection of higher quality land and desire for social presence are shown to be more important. In fact, model performance is improved when agents select farm plots further away from their failed farm plot. Farm selection strategies designed using these insights into the socio-agricultural behavior of the ancestral Pueblo significantly improved the model's accuracy and robustness.



\section*{Introduction}
Exposing the mechanics of the human decision-making processes that cause complex, society-scale phenomena is a difficult endeavor. These decision-making processes are often driven by multiple causal factors \cite{Epstein2013} with researchers having no direct means of measuring how these factors contribute to society-scale phenomena. Abductive reasoning via data-driven modeling and simulation techniques can overcome these issues by `growing' artificial societies \cite{epstein1999agent} and adjusting their configurations until adequate matches between simulation results and real world data are achieved. Agent-based modeling (ABM) in particular, offers the benefit of representing behaviors as human-interpretable rules. These rules are driven by the agent's autonomous evaluation of a variety of factors that are hypothesized by the modeler to be important in the decision-making process being modeled, indicated by the ability of ABM to simulate real world observations. However, a particular behavior rule only represents a single hypothetical decision-making process contained within a large space of possible, alternate decision-making processes. Exploring this vast space of rules requires the repeated re-implementation of multiple versions of the same ABM with different embedded decision-making processes\cite{grimm2005pattern}; this is a tedious task, involving the comparison of a massive number of combinations of causal factors. Thus, researchers often resort to modeling the most intuitive decision-making processes, if not process, which risks a subjective and inaccurate representation of the actual individual behavior\cite{grune2009explanatory}.

The current standard of ABM exploration, parameter calibration, is a black-box technique and does not perform white-box rule exploration. Parameter calibration works on the assumption of the correctness of a predefined rule and fine tunes the coefficients of its constituent factors, but cannot it easily experiment between different structures and operators through which these factors combine. 
Unless the importance of factors and how they are structured in the behavior rule are established, the underlying behavior rule merely remains an untested hypothesis of the actual individual behavior \cite{grune2009explanatory,grimm2005pattern,epstein1999agent}. Parameter calibration tools are readily available for ABM frameworks, such as BehaviorSearch \cite{stonedahl2010behaviorsearch} for NetLogo \cite{netlogo1999}, and OptQuest \cite{laguna2003optquest} for AnyLogic \cite{borshchev2013big}. Inductive games have been used to infer the decision-making of societies via game theory\cite{dedeo2010inductive}, yet no established methodology exists for ABMs. As ABM rules are implemented as program instructions, genetic programming \cite{koza1992genetic} is a highly suitable technique for model discovery. However, research into using genetic programming with ABMs for exploration of causality has been limited\cite{manson2005agent,zhong2014automatic,gunaratne2017alternate}.

To meet this need, we introduce evolutionary model discovery, a technique for agent rule exploration and causal factor importance measurement, which combines the automated program generation capability of genetic programming \cite{koza1992genetic} with the factor importance evaluation capability of random forest regression\cite{breiman2002manual,genuer2010variable,louppe2014understanding,saabas2014interpreting,saabas_2019,hutter2014efficient}. Unlike current standard techniques like pattern-oriented modeling \cite{grimm2005pattern} and model selection \cite{stratton2015automated}, evolutionary model discovery has the advantage of avoiding manual and repetitive re-implementation of models through automated program generation, resulting in a greatly reduced risk of implementation errors. Agent rules generated through genetic programming consist of functions of primitives that are easily comparable, as they follow a common representation. Using this representation, differences between candidate models are isolated to the code implementing the decision under scrutiny, to facilitate factor analysis and to avoid the need to compare two completely different implementations. The comparability of candidate models is important in drawing insights into the causes of the society-level phenomena being simulated. The stochasticity of genetic programming allows for the exploration of a vast space of possible agent rules, while selection of fitter models for breeding the next generation of rules ensures the exploitation of stronger factor interactions. Assumptions on agent behavior can be relaxed and rules with deeper factor interactions evaluated. Genetic programming, random forest training, and factor importance evaluation are all easily parallelizable techniques, which is important considering the large search space that can result even from a simple factor set.

In this study, we employ evolutionary model discovery to discover plausible interactions of causal factors by comparing their importance in the farm plot selection of the ancestral Pueblo community modeled in the Artificial Anasazi \cite{anasazinetlogo2010,dean2000understanding}. The ABM simulates the population dynamics of the Long House Valley between the years 800 AD to 1400 AD during which there was a sudden population collapse around 1350 AD. The original model demonstrated that this collapse was not caused by environmental factors alone. The model is data driven and simulations attempt to match the annual population time-series measuring households in the valley, which was estimated through data gathered from archaeological digs \cite{dean2000understanding}. The agents in the model represent households, and are dependent on the agricultural success of their farm plot for sustenance and reproduction. The farm plot selection strategy originally implemented dictates that upon depletion of a household's current farm plot, the agent moves to the next closest available plot of land. In other words, the sole factor influencing this decision is the minimization of distance over the complete set of available plots of land in the valley.

We argue that this behavior is not entirely human-like, and could have been influenced by factors other than distance. Instead, we hypothesize that nine different factors and four different social structures governing information flow may have driven the farm plot seeking behavior of the modeled Pueblo society. Specifically, we hypothesize that the following factors could have had significant importance: distance ($F_{Dist}$), dryness of the farm-land ($F_{Dry}$), quality of farm-land ($F_{Qual}$), yield of the land in the previous year ($F_{Yield}$), water availability ($F_{Water}$), social presence near the potential farm land ($F_{Soc}$), homophily by age ($F_{Age}$), homophily by agricultural success ($F_{Agri}$), and inter-zone migration ($F_{Mig}$), under the following possible social connectivity configurations: full information of the valley ($S_{All}$), information provided by family immediate family members ($S_{Fam}$), information provided by the most productive households ($S_{Perf}$), or information from the nearest neighbors ($S_{Nhbr}$). We consider the coefficients of these factors in the evolved agent behavior rules as the factors' `presence' in that particular behavior rule. Each factor's presence is then analyzed for its importance at predicting the ABM's fitness through feature importance analysis on a random forest trained on data generated by the genetic program. Utilizing a random forest for this purpose allowed us to measure both main effects of the factors' presence and the joint contributions of factors towards the ABM's fitness. After identifying the most important factors, we determined the optimal presence for them. With these insights we were able to construct causally accurate and robust farm selection procedures. 

Our results falsify the original assumption\cite{anasazinetlogo2010,dean2000understanding} that closeness was the sole causal factor governing farm plot selection of the ancestral Pueblo society. Instead, evolutionary model discovery reports the most important factors as quality, social presence, migration from zone, distance, and dryness in order of decreasing importance. In particular, the selection of higher quality land that either had a higher social presence or was located in a different zone was shown to be more likely behavior and versions of the Artificial Anasazi with these farm selection strategies were significantly more robust against random initialization of parameters. Our results indicate that the farm selection strategy was likely more human-like than that implemented in original version of the model \cite{anasazinetlogo2010,dean2000understanding}. 


\section*{Methodology}

\subsection*{Farm Plot Selection in the Artificial Anasazi}
The Artificial Anasazi is an agent-based model of the Kayenta Anasazi during the years of 800 AD to 1350 AD \cite{anasazinetlogo2010,dean2000understanding}. This model was initially developed as part of a larger effort to study the ancestral Pueblo civilization that occupied the Long House Valley region. The ABM is implemented in NetLogo \cite{netlogo1999,anasazinetlogo2010}. Archaeological excavations provide annual population time series data as estimated counts of households that existed in the valley during the period of study. Annual data on water sources and estimated soil dryness (Adjusted Palmer Drought Severity Index) for each grid location on the map are provided. The model used a normal distribution to map relative quality of soil over the map. The agent-based model simulates the rise and fall of households over a geographic map of the valley over time and produces a time series of annual household count. The original purpose of the Artificial Anasazi was to test if environmental factors could have triggered the sudden disappearance of the Anasazi from the Long House Valley around 1350 AD. 

Critics of the Artificial Anasazi have argued that the agent-based model itself is but a single candidate explanation of the social phenomenon at hand, the rise and fall of the Anasazi population over time \cite{grune2009explanatory}.  However, we view this as an advantage as the Artificial Anasazi can be used as a test-bed to discover multiple plausible explanations of the population dynamics of the Long Valley at the time. Testing combinations of hypothesized factors that may have influenced actual decision-making processes of the individuals results in a vast search space of plausible Artificial Anasazi behavior results.

We concentrated on a particular sub-model of the Artificial Anasazi: the farm plot selection strategy. The households perform farm plot selection under two conditions: 1) when a new child household is hatched by a household that has enough resources to increase its family size, or 2) when the current farm plot is unable to produce enough yield to satisfy the nutrition needs of the household anymore. The original model, hypothesizes that the households simply selected the next closest available farm plot to the household's current farm plot during farm plot selection, i.e., minimizing over distance. A patch must be free of farms or households and not be located inside a water body to be available. Consequently, the original farm selection strategy ignores other sensory data available to the households regarding the land and the state of other households in the valley.

\subsection*{Hypothesized Alternate Factors Influencing Farm Plot Selection}
Human social behavior is rarely entirely rational. Accordingly, our hypothesis proposed that the farm selection decisions of the ancestral Pueblo were complex, and took into account the state of the potential farm plots available to them and the social influences of other households around them. Agent\_Zero \cite{epstein1999agent} models the human decision making process into three dimensions: social, emotional and rational. Similarly, we defined factors that we hypothesized to influence the farm plot selection process within these dimensions. The social component is expressed through four mutually exclusive social connectivity configurations through which the agent could receive information on a subset of potential farm plots, $s$, out of the entire set of potential farm plots in the valley, $S_{All}$. The received information is then processed through a utility function $f(x)$ defined as a combination of factors and operators, $F$, which consider both the internal state of the household and the conditions of the farm plot and its surroundings in order to determine the next farm plot $x' \in $s$ \subset S_{All}$ as in Eq~(\ref{eq:farmplotselection}).
\begin{eqnarray}
\label{eq:farmplotselection}
    x' = \argmax_{x \in s \subset S_{All}}f(x)
\end{eqnarray}

Households in the original Artificial Anasazi model consider a single factor, distance, which we will refer to as $F_{Dist}$, and choose the potential farm plot with minimal distance to their current farm location. No further factors are considered in the decision making process. Furthermore, the original model assumes that the households have complete information of the valley, and every potential farm plot is compared. Therefore, the farm selection process of the original Artificial Anasazi can be represented as in Eq~(\ref{eq:farmplotselection}).
\begin{eqnarray}
\label{eq:farmplotselectionoriginal}
    x' = \argmax_{x \in S_{All}} (-F_{Dist}(x))
\end{eqnarray}

Arguing that the farm selection decision may have been more complex, considering a variety of other factors, we proposed an extended factor set consisting of four social and five rational factors, namely: homophily by age ($F_{HAge}$), homophily by agricultural productivity ($F_{HAgri}$), social presence ($F_{Soc}$), migration from current zone ($F_{Mig}$), comparison of quality ($F_{Qual}$), comparison of dryness ($F_{Dry}$), comparison of yield ($F_{Yeild}$), comparison of water availability ($F_{Water}$), and comparison of distance ($F_{Dist}$). Additionally, the numerical operators $+$ and $-$ are included in $F$, for the aggregation of sub-scores reported by the social/emotional and rational factors.

Four hypothesized configurations of social connectivity were included $F$. These configurations determined the subset of all viable farm plots that were to be considered by the households for comparison. 1) Full information ($S_{All}$): Households had complete knowledge of all potential farm plots in the valley. Full information was used by agents in the original version of the model, assuming that each household knew and compared every potential farm plot in the Long House Valley. 2) Family inherited information ($S_{Fam}$): Households solely depended on information available through their `family'. Families are defined as a household's parent household, sibling households, any surviving grandparents, and the household itself. 3) Nearest-neighbor information ($S_{Neigh}$): agents only consider the farm plots known to their neighboring households within a fixed radius of their current location. 4) Best performers $S_{Perf}$: Households only consider potential farm plots known to the best performing households, demonstrating a leadership dynamic.

Four social/emotional factors were included in $F$: two types of homophily (the tendency for social entities to congregate among those with similar traits), need for social presence, and one of fleeing/migration. Each social/emotional factor returned a sub-score representing the desirability of each evaluated farm plot. Sub-scores were normalized within the factors, to lie in the range of 0 to 1, for fair comparison. 1) Homophily by age ($F_{HAge}$): Households prefer to select farm plots near other households that are of similar age, where age is measured as the number of simulation steps the household has survived since splitting from its parent. 2) Homophily by agricultural productivity ($F_{HAgri}$): Households tend to select farm plots near other households with a similar corn stock to itself. 3) Social presence ($F_{Soc}$): Agents score potential farm plots with many nearby households higher than those in isolation. 4) Fleeing/migration ($F_{Mig}$): Agents score potential farm plots that are in a completely different zone than the current one with a full sub-score, while patches in the same zone receive a sub-score of zero.

Five Rational factors considered for the farm selection process were logical comparisons of sensory data on the potential farm plots already available to the households in the original model. Similar to the social/emotional factors, rational factors also returned a normalized sub-score of farm plot desirability between 0 and 1. 1) Comparison of quality ($F_{Qual}$): Higher sub-scores were reported for potential farm plots with higher quality of land. 2) Comparison of dryness ($F_{Dry}$): Higher sub-scores were reported for potential farm plots with higher dryness of land. 3) Comparison of yield ($F_{Yeild}$): Higher sub-scores were reported for potential farm plots that were known to have higher yield in the previous year. 4) Water availability ($F_{Water}$): Higher sub-scores were reported for potential farm plots with more nearby water sources. 5) Comparison of distance ($F_{Dist}$): Higher sub-scores were reported for potential farm plots that were closer to the current farm plot location.

\subsection*{Evolutionary model discovery}
\label{sec:emd}
Evolutionary model discovery allows agent-based modelers to explore the importance of a hypothesized set of factors affecting individual-level decision making towards a macro, society-level outcome. Accordingly, evolutionary model discovery requires the modeler to identify the particular agent behavior rule being evaluated within the original agent-based model. The modeler must also provide a set of hypothesized factors and combining operators that the modeler hypothesizes to affect the decision-making process represented by the agent behavior rule. 

A factor $F_i \in F$, where $F$ is the modeler's set of hypothetical factors and operators, is defined as in Eq~(\ref{eq:causalfactor}). Where $C$ is the set of commands defined within $F$ that are applied on the $n$ number of input parameters $P$ to produce an output return value $R$, where the type of each parameter $t_{P_j}$ and the type of the return value $t_R$ are each an element of the set $T$ of all possible parameter and return types defined by the modeler. A factor is considered an operator if $C$ resembles an operation on one or more factors, which it accepts as parameters, rather than resembling a decision-making step. In order for a factor or operator $F_i$ to accept another $F_j$ as an input, the condition Eq~(\ref{eq:factorcombinationcondition}) must be met. 
\begin{eqnarray}
\label{eq:causalfactor}
    F_i = (C,R,P \quad| \quad t_R, t_{P_k} \in T \quad \forall k = 1...n )
\end{eqnarray}
\begin{eqnarray}
\label{eq:factorcombinationcondition}
    \exists k,\; t_{R_{f_i}} = t_{P_{{f_j},k}}
\end{eqnarray}

An agent behavior rule $b \in B$ is represented as a tree of factors combined under this condition. Depending on $T$ and the factor definitions, the space of behavior rules $B$ can be infinitely large. To prevent the construction of such undesirably large trees, we specify a maximum depth for all $b$. There must be at least one $F_i$ of which $t_{R_{F_i}}$ is the return type expected by the entire agent behavior rule.

Given the ABM and $F$, evolutionary model discovery performs two stages of analysis. First, models driven by alternate decision making processes consisting of combinations of elements of $F$ are evolved through genetic programming \cite{koza1992genetic,langdon2014optimizing,petke2017genetic}. Genetic programming performs automated program implementation and is a suitable approach towards automating the rule discovery process \cite{manson2005agent,manson2006bounded, zhong2014automatic, gunaratne2017alternate}.
Genetic programming evolves generations of programs through crossover and mutation operators performed on a representation consisting of primitives and terminals that combine to define program statements. Primitives are defined as a set of functions that encode program statements and may be strongly typed to only accept child and parent primitives that are compatible with the arguments and return statements accepted by its program statement. Primitives with no arguments are considered terminals. The syntax tree representation is perhaps the most common representation used in genetic programming, and arranges the primitives and terminals into a tree structure, a representation compatible with $b$. Programs in a generation that have a closer fit to data are more likely to be selected for reproduction through crossover and mutation to populate the next generation of programs.

Second, factor and factor interaction importance was assessed by random forest feature importance measurement. A random forest regressor was trained on the factor presence to fitness data produced by the genetic program. Random forests are an ensemble learning algorithm consisting of a forest of randomized decision trees \cite{breiman2002manual,genuer2010variable,louppe2014understanding}. The two most common factor importance measurement techniques for random forests are gini importance (or mean decrease in impurity), and permutation importance (or mean decrease in accuracy)\cite{breiman2002manual,genuer2010variable,louppe2014understanding}. However, both gini importance and permutation importance are unable to quantify the importance of factor interactions, as they consider the global importance each factor has for the random forest. Functional analysis of variance \cite{hutter2014efficient, dang2018analysis} is able to quantify the importance of factor interactions, yet lacked precision considering the inherent heteroskedasticity of the data produced by the genetic program, caused by its tendency to explore and test models of higher fitness. Instead, joint contribution \cite{saabas2014interpreting,saabas_2019} was used for this purpose as it has been successfully used to assess the importance of variable interactions in a large number of recent studies \cite{lundberg2018consistent, smith2017forecasting, Beillevaire2018, rea2018initial, granetz2018machine, morice2018learning, bastrakova2017improving}.

Twenty genetic programming runs were executed with the objective of minimizing the (RMSE) between the simulated household count to the actual household count over 550 simulation ticks of the Artificial Anasazi. Details on the RMSE calculation can be found in \cite{gunaratne2017alternate}. In order to ensure robustness of the evolved rules, the parameters of the ABM were randomly initialized with values $\pm5\%$ about the optimal parameter values found through Stonedahl's calibration of the Artificial Anazasi through a genetic algorithm \cite{stonedahl2010behaviorsearch} (ie: water source distance = $(10.925, 12.075)$, death age span = $(9.5, 10.5)$, min fertility = $(0.1615, 0.1785)$, base nutrition need = $(175.75, 194.25)$, fertility span = $(0.0285, 0.0315)$, min fertility ends age = $(27.55, 30.45)$, harvest variance = $(0.418, 0.462)$, harvest adjustment = $(0.608, 0.672)$, maize gift to child = $(0.4465, 0.4935)$, min death age = $(38.0, 42.0)$, fertility ends age span = $(4.75, 5.25)$). The genetic program was implemented with the Distributed Evolutionary Algorithms in Python library (DEAP)\cite{DEAP_JMLR2012} and parallelized by SCOOP\cite{SCOOP_XSEDE2014}. Each genetic program run was executed for 100 generations over populations of 50 individuals. Syntax trees of minimum depth 4 and maximum depth 10 were used to avoid trees exhibiting bloat. The Half-and-Half tree builder was used for initialization \cite{koza1992genetic}.  To accommodate the high computational cost, the genetic program runs were distributed across a 48 vcpu Amazon Web Services EC2 instance. The random forest and gini importance algorithm of Scikit-learn \cite{scikit-learn} were used, while ELI5\cite{korobov_lopuhin_2019} was used for permutation accuracy importance, and tree interpreter\cite{saabas_2019} for joint contribution measurement.

Finally, new farm selection strategies were designed taking into account the insights gained through evolutionary model discovery. The robustness of the Artificial Anasazi with these new strategies were tested against the original model by comparing the RMSE of 100 runs of each model under randomized initialization of parameters within the ranges above.


\section*{Results}

\begin{figure}[!h]
\centering
\includegraphics[width=0.6\linewidth]{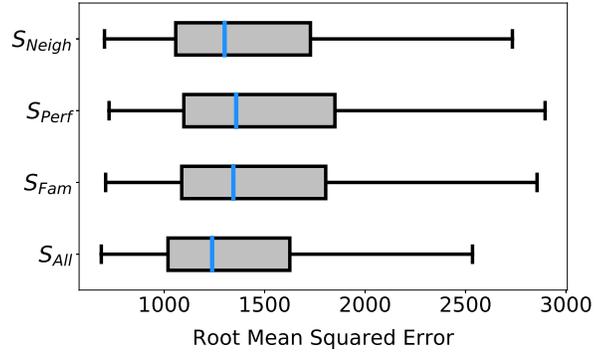}
\caption{{\bf Best fit to data was obtained under S\textsubscript{All}.} Comparison of the RMSE produced by the Artificial Anasazi model when agents had full information ($S_{All}$), information through family households ($S_{Fam}$), information through the households with most agricultural success ($S_{Perf}$), or information through neighboring households ($S_{Neigh}$). Models that used $S_{All}$ produced the lowest RMSE overall $\argmax_{x \in S_{All}} f(x) <  \argmax_{x \in S_{Fam}} f(x)$ ($p = 2.045 \times 10^{-113}$), $\argmax_{x \in S_{All}} f(x) < \argmax_{x \in S_{Neigh}} f(x)$ ($p=4.856 \times 10^{-154}$), $\argmax_{x \in S_{All}} f(x) < \argmax_{x \in S_{Perf}} f(x)$ ($p = 1.983 \times 10^{-57}$).}
\label{fig:S_RMSE}
\end{figure}

The resulting best farm selection strategies evolved by the genetic program by run are provided in Table \ref{tab:20GPRunsCode} along with their respective RMSE values. 15 of the runs produced RMSE values lower than the current best RMSE in the literature obtained through parameter calibration of the Artificial Anasazi model with the original farm plot selection by closeness (733.6) \cite{stonedahl2010evolutionary}. All best scoring rules for each run utilized $S_{All}$, i.e., the model produced best results when the agents had full information regarding available farm plots as shown in Fig. \ref{fig:S_RMSE}, comparing $S_{All}$, $S_{Fam}$, $S_{Neigh}$, and $S_{Perf}$ over the complete factor presence to fitness data. One-tailed Mann-Whitney U tests comparing the fitness of all rules by their social connectivity configurations confirmed that rules with $S_{All}$ had significantly ($\alpha=0.05$) lower RMSE than the other three configurations: $\argmax_{x \in S_{All}} f(x) <  \argmax_{x \in S_{Fam}} f(x)$ ($p = 2.045 \times 10^{-113}$), $\argmax_{x \in S_{All}} f(x) < \argmax_{x \in S_{Neigh}} f(x)$ ($p = 4.856 \times 10^{-154}$), $\argmax_{x \in S_{All}} f(x) < \argmax_{x \in S_{Perf}} f(x)$ ($p = 1.983 \times 10^{-57}$). Also, rules with $S_{Neigh}$ were shown to have significantly (alpha=0.05) lower RMSE than those with $S_{Fam}$ and $S_{Perf}$: $\argmax_{x \in S_{Neigh}} f(x) < \argmax_{x \in S_{Fam}} f(x)$ ($p = 3.535 \times 10^{-14}$), $\argmax_{x \in S_{Neigh}} f(x) < \argmax_{x \in S_{Perf}} f(x)$ ($p = 2.339 \time 10^{-24}$). Finally, rules with $ S_{Fam}$ were shown to have significantly (alpha=0.05) lower RMSE than rules with $S_{Perf}$: $\argmax_{x \in S_{Fam}} f(x) < \argmax_{x \in S_{Perf}} f(x)$ ($p = 0.012$). Accordingly, the rest of the analyses detailed in this paper were performed on rules where the social connectivity configuration was $S_{All}$. 

\begin{table}[tbh!]
\begin{adjustwidth}{-2.25in}{0in} 
    \centering
    \caption{\bf The candidate farm selection strategies of models produced by the evolutionary model discovery process along with their best fitness as reported by the genetic programming search.}
    \begin{tabular}{c|l|c}
        \hline
        GP Run&Best scoring rule&Best Fitness\\\hline
        0&$\displaystyle \argmax_{x \in S_{All}} (F_{Mig}(x))$& 753.430820 \\\hline
        1&$\displaystyle \argmax_{x \in S_{All}} (-F_{Dist}(x)-F_{Dry}(x)+2*F_{Mig}(x))$& 755.270812 \\\hline
        2&$\displaystyle \argmax_{x \in S_{All}} (F_{Yield}(x)+F_{HAgri}(x))$& 709.502643 \\\hline
        3&$\displaystyle \argmax_{x \in S_{All}} (F_{Mig}(x)-F_{HAgri}(x))$& 738.949931 \\\hline
        4&$\displaystyle \argmax_{x \in S_{All}} (F_{Mig}(x))$& 730.475188 \\\hline
        5&$\displaystyle \argmax_{x \in S_{All}} (F_{Dist}(x))$& 752.519767 \\\hline
        6&$\displaystyle \argmax_{x \in S_{All}} (F_{Dist}(x))$& 728.293210 \\\hline
        7&$\displaystyle \argmax_{x \in S_{All}} (F_{Yield}(x))$& 714.205153 \\\hline
        8&$\displaystyle \argmax_{x \in S_{All}} (F_{Dist}(x)-F_{Dry}(x))$& 734.249957 \\\hline
        9&$\displaystyle \argmax_{x \in S_{All}} (4*F_{Dist}(x)+F_{Dry}(x)+F_{Qual}(x)+F_{Water}(x)+F_{Soc}(x)+F_{HAge}(x))$& 701.208243 \\\hline
        10&$\displaystyle \argmax_{x \in S_{All}} (F_{Dist}(x)+F_{Qual}(x)+F_{Water}(x)-F_{Yield}(x)+F_{Mig}(x)+F_{Soc}(x))$& 720.281195 \\\hline
        11&$\displaystyle \argmax_{x \in S_{All}} (F_{Mig}(x))$& 723.633194 \\\hline
        12&$\displaystyle \argmax_{x \in S_{All}} (F_{Dist}(x)+F_{Qual}(x)+2*F_{Yield}(x)+2*F_{Mig}(x)+F_{Soc}(x)+F_{HAgri}(x))$& 687.122260 \\\hline
        13&$\displaystyle \argmax_{x \in S_{All}} (F_{Dist}(x)+F_{Soc}(x))$& 732.189183 \\\hline
        14&$\displaystyle \argmax_{x \in S_{All}} (F_{Qual}(x))$& 728.772255 \\\hline
        15&$\displaystyle \argmax_{x \in S_{All}} (F_{Qual}(x))$& 706.282521 \\\hline
        16&$\displaystyle \argmax_{x \in S_{All}} (F_{Dist}(x)+2*F_{Qual}(x)+F_{Yield}(x)+F_{Soc}(x)+3*F_{HAge}(x))$& 715.957401 \\\hline
        17&$\displaystyle \argmax_{x \in S_{All}} (F_{Mig}(x))$& 715.468378 \\\hline
        18&$\displaystyle \argmax_{x \in S_{All}} (-F_{Dist}(x)+F_{Soc}(x)-F_{HAgri}(x))$& 701.438522 \\\hline
        19&$\displaystyle \argmax_{x \in S_{All}} (F_{Qual}(x)+F_{Mig}(x)+F_{Soc}(x))$& 701.300934 \\\hline
    \end{tabular}
    \label{tab:20GPRunsCode}
    \end{adjustwidth}
\end{table}

\begin{figure}[!h]
\includegraphics[width=\linewidth]{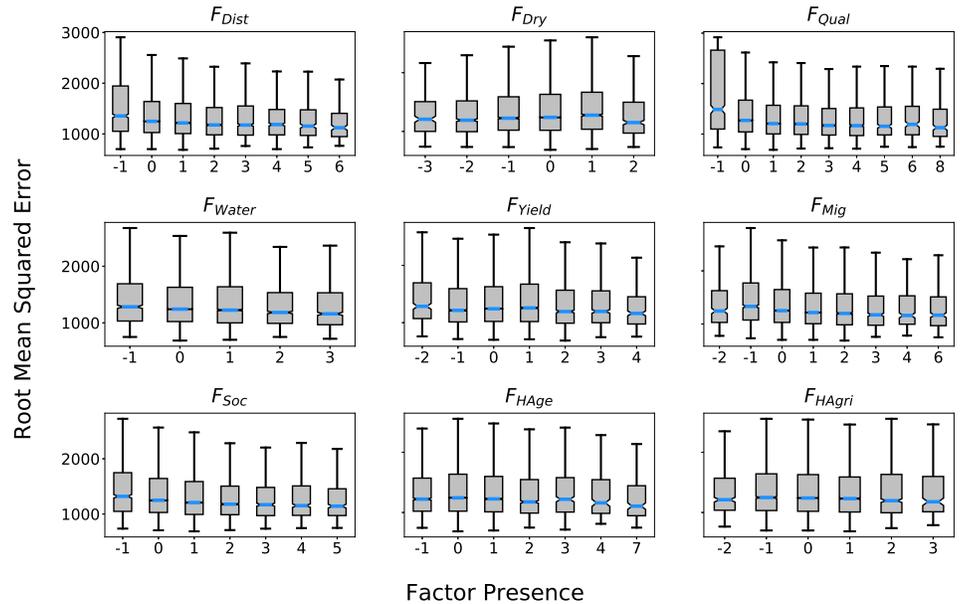}
\caption{{\bf RMSE vs Factor Presence under S\textsubscript{All}.} RMSE distributions by factor presence produced by evolutionary model discovery of the farm selection strategy of the Artificial Anasazi under $S_{All}$. Only presence values that appeared at least 200 times in the genetic program are displayed. Most factors display negative correlations to RMSE, while $F_{Dry}$ shows a positive correlation.}
\label{fig:CoefFactorMSE}
\end{figure}

Fig. \ref{fig:CoefFactorMSE} displays the distribution of RMSE against factor presence, for presence values that were recorded in at least 200 rules across the 20 genetic program runs. Negative correlations to RMSE (higher fitness) are seen between $F_{Dist}$, $F_{Qual}$, $F_{Water}$, $F_{Yield}$, $F_{Mig}$, $F_{Soc}$, and $F_{Age}$, and in general the genetic program favored the positive presence of these factors, and evolved more rules with these factors having a positive effect on farm selection. $F_{Dry}$ on the other hand had a negative correlation to RMSE for presence less than 2.

\begin{figure}[!h]
\centering
\includegraphics[width=0.8\linewidth]{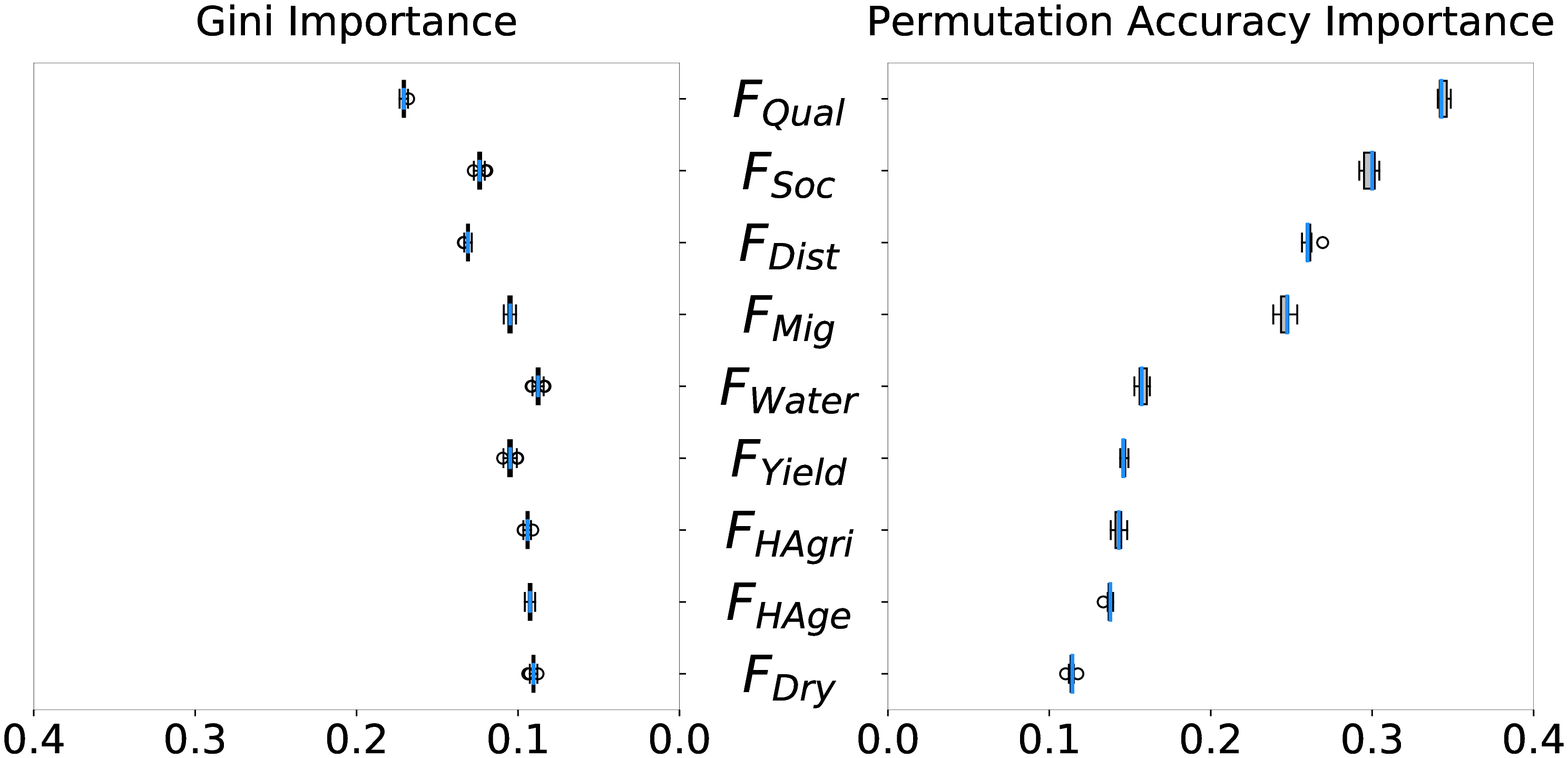}
\caption{{\bf $F_{Qual}$, $F_{Soc}$, $F_{Dist}$, and $F_{Mig}$ have highest Gini and Permutation Accuracy Importance.} Gini importance and permutation accuracy importance of the hypothesized factors towards a random forest's ability to predict the models' RMSE. Gini importance results are less decisive than permutation accuracy importance. Both techniques agree that $F_{Qual}$, $F_{Soc}$, $F_{Dist}$, and $F_{Mig}$ are the most important factors.}
\label{fig:Main_Effects}
\end{figure}

\begin{figure}[!h]
\centering
\includegraphics[width=0.8\linewidth]{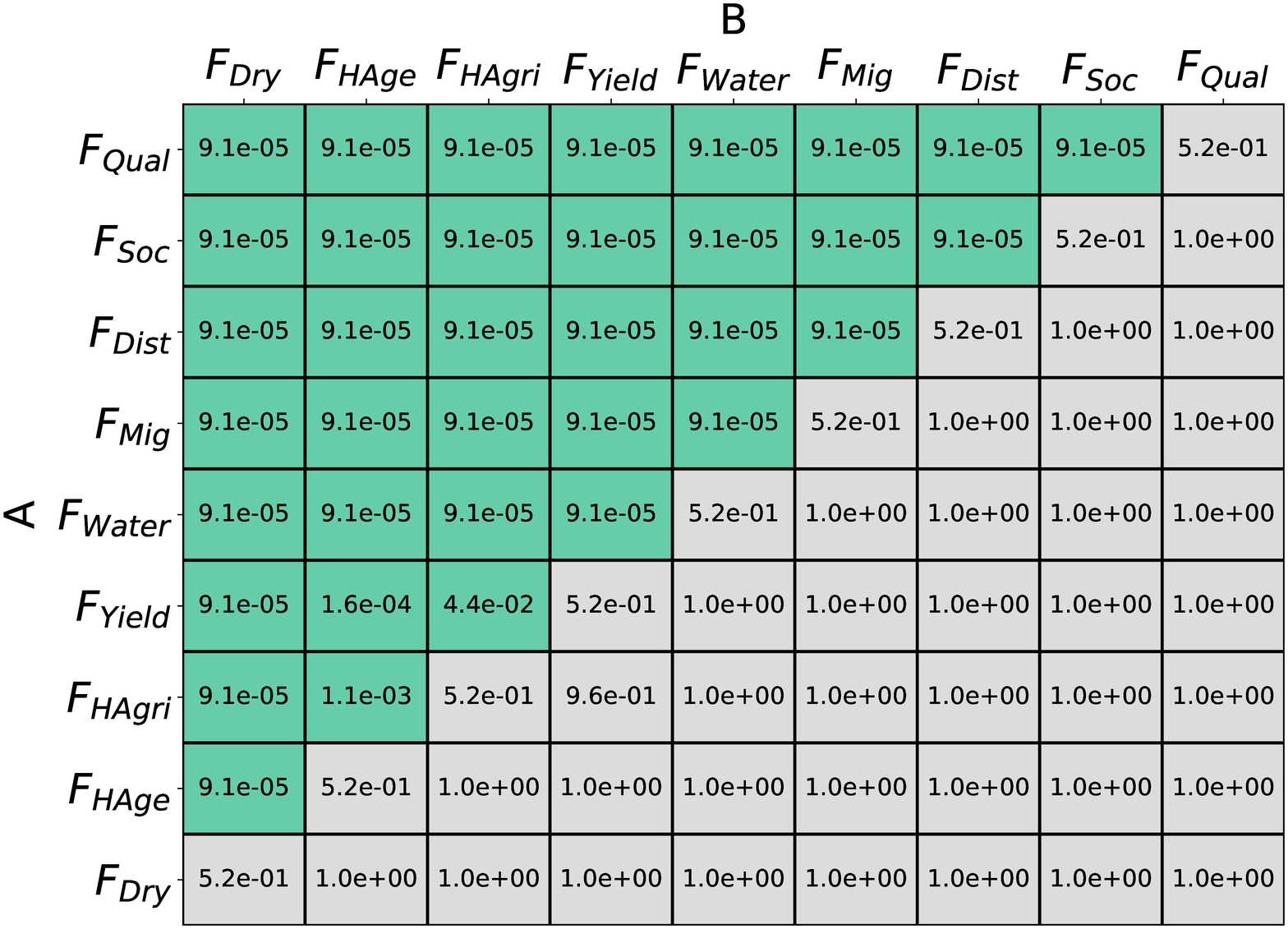}
\caption{{\bf {Statistical confirmation of the existence of order by importance among causal factors.}} Results from systematic Mann-Whitney U tests on the permutation accuracy importance results. The cells contain p-values for the alternate hypothesis that $A > B$ (null hypothesis $A = B$). Green cells indicate agreement of the alternate hypothesis. The results indicate a clear ordering of the factors by importance.}
\label{fig:MannWhitneyUMainEffects}
\end{figure}

The random forest fit the factor presence to fitness data best for a forest of 520 regression trees, testing from 10 to 1000 trees with a train/test split 90\%-10\%. Accordingly, a forest of 520 trees was used for factor importance determination. Factor importance under $S_{All}$ obtained through both the gini importance and permutation accuracy importance techniques can be seen in Fig. \ref{fig:Main_Effects}. Gini importance generally had less precise estimations than permutation accuracy importance. Yet both techniques indicated $F_{Qual}$ as the factor of highest importance towards RMSE prediction. $F_{Soc}$, $F_{Mig}$, and $F_{Dist}$ also scored higher importance values than the other factors hypothesized. Fig. \ref{fig:MannWhitneyUMainEffects} displays the p-values of one-tailed Mann Whitney U tests (alpha=0.05), comparing the permutation importance of each factor $A$ against every other factor $B$, testing the alternate hypothesis: importance of $A$ > importance of $B$. According to the results, 7 of the 9 factors showed significant difference and could be ordered in terms of permutation accuracy importance as $F_{Qual}$, $F_{Soc}$, $F_{Dist}$, $F_{Mig}$, $F_{Water}$, $F_{Yield}$, $F_{HAgri}$, $F_{HAge}$, and $F_{Dry}$.

\begin{figure}[!h]
\centering
\includegraphics[width=0.8\textwidth]{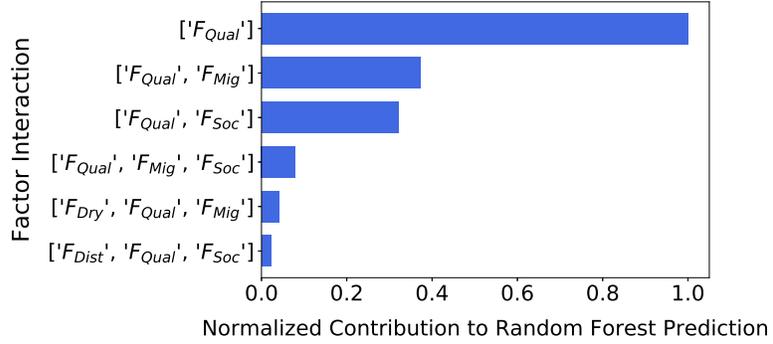}
\caption{{\bf {F\textsubscript{Qual}, [F\textsubscript{Qual},F\textsubscript{Mig}], and [F\textsubscript{Qual}F\textsubscript{Soc}] have highest joint contribution to farm plot selection.}} Ordered barchart of highest normalized joint contribution scores of factors and interactions of three or less under $S_{All}$. Again, $F_{Qual}$ shows a far larger contribution to the random forest's ability to predict model RMSE than other factors and factor interactions, and is present in all of the highest contributing interactions. Interactions [$F_{Qual}$,$F_{Mig}$] and [$F_{Qual}$,$F_{Soc}$] also demonstrate high joint contribution.}
\label{fig:Joint_Contributions}
\end{figure}

Fig. \ref{fig:Joint_Contributions} compares the top ten joint contributions towards RMSE prediction of the random forest by individual factors, and joint contributions of factors considered in pairs and triples. Again, $F_{Qual}$ demonstrates far higher importance than any other factor or factor interaction. The factor pairs $(F_{Qual},F_{Mig})$ and $(F_{Qual},F_{Soc})$ also demonstrate high importance, followed by $(F_{Qual},F_{Mig},F_{Soc})$, $(F_{Dry},F_{Qual},F_{Mig})$, and $(F_{Dist},F_{Qual},F_{Soc})$. Overall, $F_Qual$ is present in all highest scoring joint contributions. Despite $F_{Dry}$ having very low individual importance, $F_{Dry}$ showed higher importance when considered in combination with $F_{Qual}$ and $F_{Mig}$.

\begin{figure}[!h]
    \centering
    \includegraphics[width=\linewidth]{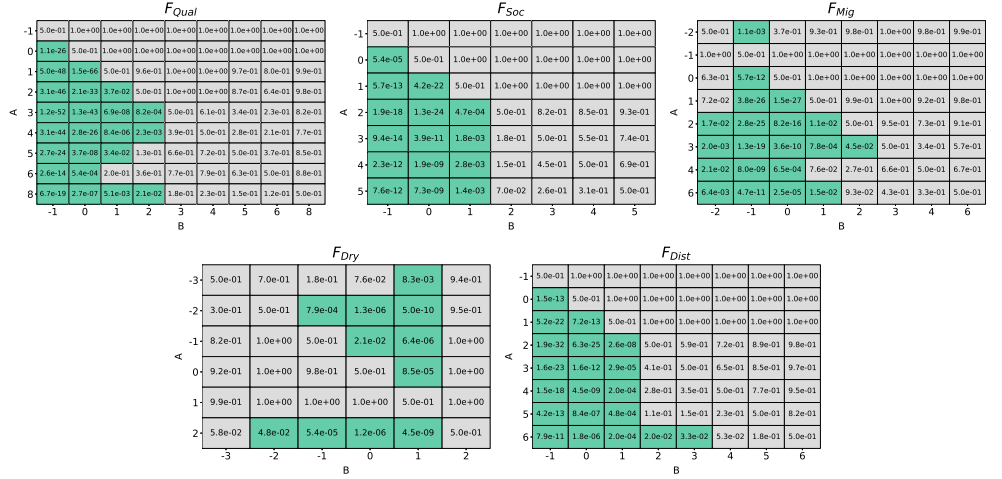}
    \caption{{\bf Optimal presence scores for causal factors with highest importance.} Results from systematic one-tailed Mann-Whitney U tests between presence values of the 5 most important factors for the alternate hypothesis: $\textrm{RMSE for presence }A < \textrm{RMSE for presence }B$ (null hypothesis: $\textrm{RMSE for presence }A = \textrm{RMSE for presence }B$) for $\alpha=0.05$. Green cells indicate agreement of the alternate hypothesis. Results indicate that for $F_{Qual}$, $F_{Soc}$, $F_{Mig}$, and $F_{Dist}$ RMSE is generally lower for higher, positive presence. For $F_{Dry}$, both negative and higher positive presence may provide low RMSE scores.}
    \label{fig:MannWhitneyU_Top5Factors}
\end{figure}

Considering the evidence of $F_{Qual}$, $F_{Soc}$, $F_{Mig}$, $F_{Dist}$, and $F_{Dry}$ as important factors, Fig. \ref{fig:MannWhitneyU_Top5Factors} demonstrates Mann Whitney U tests conducted for each factor $F_i$, for the alternate hypothesis that RMSE when presence of $F_i$ was $A$, is less than the RMSE when presence of $F_i$ was $B$ in rules with $S_{All}$. Models with positive presence of $F_{Qual}$, $F_{Soc}$, $F_{Dist}$, and $F_{Mig}$ showed significantly higher fitness (with the exception of when presence of $F_{Mig}$ = -2). Models with strong positive or negative presence of $F_{Dry}$ showed lower RMSE overall, most likely a result of $F_{Dry}$'s interaction with $F_{Qual}$, $F_{Soc}$, or $F_{Mig}$. The lowest median RMSE for $(F_{Qual},F_{Soc})$ was 985 at presence of $F_{Soc}$ at 5 and presence of $F_{Qual}$ at; the lowest median RMSE for $(F_{Qual},F_{Mig})$ was 997 at presence of $F_{Mig}$ at 3 and presence of $F_{Qual}$ at 5. 

\begin{figure}[!h]
    \centering
    \includegraphics[width=\linewidth]{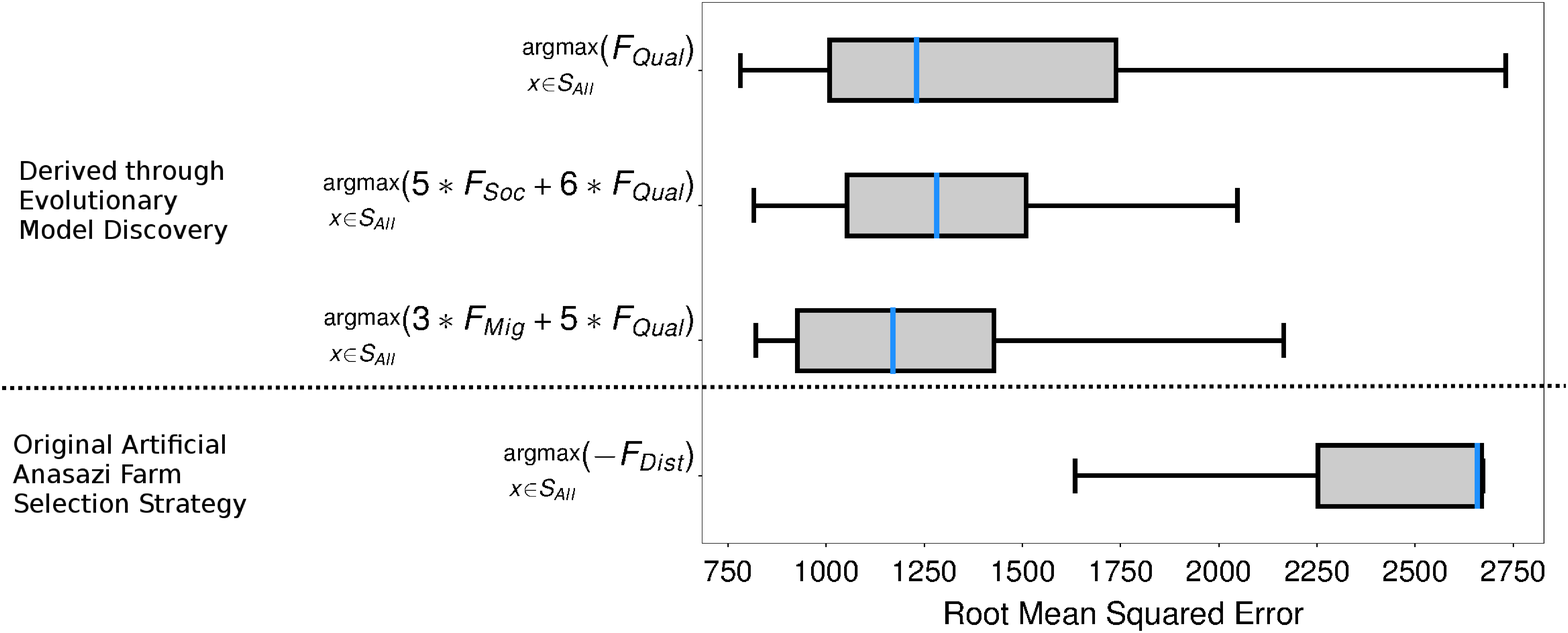}
    \caption{{\bf Models designed through evolutionary model discovery insights are significantly more robust.} Comparison between the RMSE of 100 runs of three models with farm selection strategies designed taking into consideration the insights from evolutionary model discovery, 1) $\argmax_{x \in S_{All}}(F_{Qual}(x))$,  2) $\argmax_{x \in S_{All}}(5F_{Soc}(x) + 6F_{Qual}(x))$, and 3) $\argmax_{x \in S_{All}}(3F_{Mig}(x) + 5F_{Qual}(x))$, against 100 runs of the original farm selection strategy  $\argmax_{x \in S_{All}}(-F_{Dist}(x))$ in~\cite{anasazinetlogo2010,dean2000understanding,janssen2009understanding,stonedahl2010evolutionary}, under random initialization of parameters. The three farm selection strategies derived from evolutionary model discovery are far more robust under random parameter initialization and show significantly better RMSE scores compared to the original model.}
    \label{fig:BestVsOriginal100Runs}
\end{figure}

Finally, rules following the three highest joint contributions were constructed using the best values for each factor concerned: $\argmax_{x \in S_{All}}(F_{Qual}(x))$,  $\argmax_{x \in S_{All}}(5F_{Soc}(x) + 6F_{Qual}(x))$, and $\argmax_{x \in S_{All}}(3F_{Mig}(x) + 5F_{Qual}(x))$, and RMSE was compared against the original farm selection strategy  $\argmax_{x \in S_{All}}(-F_{Dist}(x))$ for 100 runs each under random initialization of parameters within the ranges specified in section \ref{sec:emd}. Fig. \ref{fig:BestVsOriginal100Runs} shows that all three of these rules derived through evolutionary model discovery have significantly lower RMSE than that of the original farm selection strategy under randomized parameter initialization.

\section*{Discussion and Conclusion}
Despite being an excellent tool for the construction and analysis of human-interpretable explanations of social phenomena, ABMs risk premature assumptions when modeling individuals' decision-making processes. Parameter calibration alone cannot adequately explore the causal factors and their possible interactions in order to infer more accurate decision-making processes. This is primarily due to the absence of a systematic method for behavior inference and discovery. We address this issue with the introduction of evolutionary model discovery, which is able to distinguish, out of a hypothesized set, the causal factors that are important to simulate the behavior of interest. By combining automated program generation of genetic programming with feature importance evaluation of random forests, evolutionary model discovery is able to quantify the importance of these factors to the decision-making process that result in society-level phenomena simulated by the ABM. This allows for the construction of agent rules that more accurately represent the actual decision-making process of individuals and result in more robust models.

Applying evolutionary model discovery on the Artificial Anasazi we show that the socio-agricultural behavior of the ancestral Pueblo of the Long House Valley was more deliberative and informed than originally assumed. Our results indicate that, contrary to the original farm selection behavior, where households would select the next closest possible plot of land once their present farm was depleted, the households most likely selected potential farming land with higher soil quality ($F_{Qual}$). Further, it was highly likely that the households had good knowledge of the potential arable land throughout the valley, since $S_{All}$ was the best social connectivity configuration for information spread. Also, the desire to congregate into communities was indicated, as positive desire for social presence ($F_{Soc}$) was the second most important factor, and acting on information on arable land known to neighboring households ($S_{Neigh}$) was the second most successful social connectivity configuration. Further, instead of choosing closer potential farm plots ($-F_{Dist}$), choosing farm plots that were further away from the households current farm plot ($F_{Dist}$) or moving to a completely different zone in the region ($F_{Mig}$) was found to be a more likely behavior. Finally, versions of the Artificial Anasazi where farm plot selection was driven by seeking higher quality land, higher quality land with more social presence, and higher quality land in different zones, all proved to be significantly more robust than the decision to move to the next closest available plot of land (Fig. \ref{fig:BestVsOriginal100Runs}).

\section*{Supporting information}

\paragraph*{S1 File.}
\label{S1_File}
{\bf EvolutionaryModelDiscoveryArtificialAnasazi.zip}  This archive contains the Evolutionary Model Discovery Python source code. This Python package is also being actively maintained at: \href{https://github.com/chathika/evolutionarymodeldiscovery}{https://github.com/chathika/evolutionarymodeldiscovery} and documentation is available at \href{https://evolutionarymodeldiscovery.readthedocs.io/en/latest/}{https://evolutionarymodeldiscovery.readthedocs.io/en/latest/}. NetLogo models of the Artificial Anasazi with the best .

\paragraph*{S1 Table.}
\label{S1_Table}
{\bf FactorScores.csv} Factor presence to model fitness data produced by the 20 genetic programming runs on the Artificial Anasazi.

\section*{Acknowledgments}
This research was supported by DARPA SocialSim program HR001117S0018 and Amazon AWS Research Grant.
\nolinenumbers

%
%
%

\end{document}